\documentclass[twoside,fleqn]{article}
\usepackage{epsfig}
\usepackage{espcrc2}
\newcommand{\AmS}{{\protect\the\textfont2
  A\kern-.1667em\lower.5ex\hbox{M}\kern-.125emS}}

\def\H{H\hskip-8.5pt/\hskip2pt}

\def\VEV#1{\left\langle #1\right\rangle}

\def\lsim{\mathrel{\mathpalette\@versim<}}

\def\gsim{\mathrel{\mathpalette\@versim>}}

\title{CPT and Quantum Mechanics Tests with Kaons: Theory}

\author{Nick E. Mavromatos\address{King's College London, Department of
Physics, Strand, London WC2R 2LS, United Kingdom}}%
\begin{document}

\begin{abstract}
In this talk I review theoretical motivations
for possible CPT Violation and deviations from ordinary quantum
mechanical behavior of field theoretic systems in 
some quantum gravity models, and I describe the relevant 
precision tests using neutral and charged
Kaons. I emphasize
the possibly unique r\^ole of neutral-meson
factories in providing specific tests of models in which the CPT
operator is not well-defined, leading to
modifications of Einstein-Podolsky-Rosen (EPR) particle correlators.

\end{abstract}

\maketitle

\vspace*{-0.7cm}
\section{CPT symmetry and Quantum Gravity}

Any quantum theory, formulated on flat space times, is symmetric
under the combined action of CPT transformations, provided the
theory respects (i) Locality, (ii) Unitarity (i.e. conservation of
probability) and (iii) Lorentz invariance. This is the famous CPT
theorem~\cite{cpt}. An extension of this theorem to quantum gravity
is by no means an obvious one; there may be information loss, in
certain space-time foam backgrounds~\cite{hawking}, implying an
evolution from pure to mixed quantum states, and hence
decoherence~\cite{hawking,ehns}. In such  situations {\it particle
phenomenology} has to be reformulated~\cite{ehns,poland} by viewing
our low-energy world as an open quantum system.

In such cases the \$ matrix is {\it not invertible},
and this implies~\cite{wald} that the
CPT operator itself is \emph{not well-defined}, at least from an
effective field theory point of view. This is a strong form of CPT
violation. This form of CPT
Violation (CPTV) introduces a fundamental arrow of time/microscopic time
irreversibility, unrelated in principle to CP properties.
Within the
scope of the present talk I will restrict myself to decoherence
and CPT invariance tests within neutral
Kaons~\cite{ehns,lopez,huet,benatti}. As I will argue later on,
this type of (decoherence-induced) CPT Violation (CPTV)
exhibits some fairly unique effects in $\phi$ ($B$-meson, ...)
factories~\cite{bmp}, associated with a potential modification of
the Einstein-Podolsky-Rosen (EPR) correlations of the entangled
neutral Kaon ($B$-meson, ...) states produced after the decay of the
$\phi$-(or $\Upsilon$-, ...) meson.

Another fundamental reason for CPT violation (CPTV) in quantum
gravity is {\it spontaneous breaking of Lorentz symmetry
(SBL)}~\cite{sme}, without necessarily implying decoherence. In this
case the ground state of the field theoretic system is characterized
by non trivial vacuum expectation values of certain tensorial
quantities, $\langle {\cal A}_\mu \rangle \ne 0~,
~ {\rm or} \quad ~\langle {\cal B}_{\mu_1\mu_2\dots}\rangle \ne
0~$.
In this talk I  will restrict myself
to Lorentz tests using neutral Kaons~\cite{sme}.

I must stress at this point that QG-decoherence and Lorentz
Violation (LV) are in principle independent~\cite{poland}. The
important difference of CPT violation in SBL models of quantum
gravity from that in space-time foam situations lies on the fact
that in the former case the CPT operator is well defined, but it
\emph{does not commute} with the effective Hamiltonian of the matter
system. In such cases one may parametrize the Lorentz and/or CPT
breaking terms by local field theory operators in the effective
lagrangian, leading to a construction known as the ``standard model
extension'' (SME)~\cite{sme}, which is a framework to study
precision tests of such effects.

In certain circumstances one may also violate locality,
but I will not discuss this case
explicitly here. Of course violations of locality could also be
tested with high precision by means of a study of discrete
symmetries in meson systems.

I must stress that the phenomenology of CPT violation is
complicated, and there seems \emph{not} to be a \emph{single} figure
of merit for it. Depending on the precise way by which
CPT violation is realized in a
given class of models of QG, there are different ways by which we
can test the violation~\cite{poland}. I stress that within the above
frameworks, CPT violation does \emph{not necessarily} imply mass
differences between particles and antiparticles.

\vspace*{-0.4cm}
\section{Lorentz Violation and Neutral
Kaons}
\vspace*{-0.2cm}
I commence my discussion
by a very brief description of experimental tests of Lorentz
symmetry, within the SME framework, using neutral Kaons, both
single~\cite{sme} and entangled states in a $\phi$
factory~\cite{adidomenico}. In order to isolate
the terms in SME effective Hamiltonian that are
pertinent to neutral Kaon tests, one should notice~\cite{sme}
that
the relevant CPTV and LV parameter $\delta_K$ must be flavour diagonal,
C violating but P,T preserving, as
a consequence of strong interaction properties in neutral meson
evolution.
This implies that $\delta_K $ is
sensitive only to the $-a_\mu^q {\overline q} \gamma_\mu q$ quark
terms in SME~\cite{sme},
where $a_\mu$ is a Lorentz and CPT violating parameter,
with dimensions of energy, and $q$ denote quark fields, with the meson
composition being denoted by $M = q_1{\overline q}_2$.

The analysis
of \cite{sme}, then, leads to the following relation of the
Lorentz and CPT violating parameter $a_\mu$ to the CPT violating
parameter $\delta_K$ of the neutral Kaon system:
$\delta_K \simeq i{\rm sin}\widehat{\phi} {\rm
exp}(i\widehat{\phi}) \gamma \left(\Delta a_0 - {\vec \beta_K}\cdot
\Delta {\vec a}\right)/\Delta m,$ with the short-hand
notation $S$=short-lived, $L$=long-lived,
$\Delta m = m_L - m_S$, $\Delta \Gamma = \Gamma_S - \Gamma_L$,
$\widehat \phi = {\rm arc}{\rm tan}(2\Delta m / \Delta\Gamma), \quad
\Delta a_\mu \equiv a_\mu^{q_2} - a_\mu^{q_1}$, and $\beta_K^\mu =
\gamma (1, {\vec \beta}_K)$ the 4-velocity of the boosted~ kaon.
The experimental bounds of $a_\mu$ in neutral-Kaon
experiments are based on searches of sidereal variations of
$\delta_K$ (day-night effects).
From KTeV experiment~\cite{ktev} the following bounds of the $X$ and
$Y$ components of the $a_\mu$ parameter have been obtained
$\Delta a
_X, \Delta a _Y < 9.2 \times 10^{-22}~ {\rm GeV}$, where $X,Y,Z$
denote sidereal coordinates.
Complementary measurements for the $a_Z$ component can come from
$\phi$ factories~\cite{adidomenico}.

%\vspace*{-0.3cm}
\section{Quantum Gravity Decoherence and  Neutral Kaons}

QG may induce decoherence and oscillations $K^0 \leftrightarrow {\overline
K}^0$~\cite{ehns,lopez}, thereby implying a two-level quantum
mechanical system interacting with a QG ``environment''. Upon the
general assumptions of average energy conservation and monotonic
entropy increase, and the specific (to the Kaon
system) assumption about the respect of the
$\Delta S=\Delta Q$ rule by the QG medium,
the modified evolution equation for the respective density matrices
of neutral Kaon matter reads~\cite{ehns}:
$$\partial_t \rho = i[\rho, H] + \delta\H \rho~,$$
where $H$ denotes the hamiltonian of the Kaon system, that may contain
(possible) CPTV differences of masses and widths between particles and
antiparticles~\cite{lopez}, and the decoherence matrix $\delta\H $
is given by~\cite{ehns}:
$$ {\scriptsize {\delta\H}_{\alpha\beta} =\left( \begin{array}{cccc}
 0  &  0 & 0 & 0 \\
 0  &  0 & 0 & 0 \\
 0  &  0 & -2\alpha  & -2\beta \\
 0  &  0 & -2\beta & -2\gamma \end{array}\right)~.}$$
Positivity of $\rho$ requires: $\alpha, \gamma  > 0,\quad
\alpha\gamma>\beta^2$. Notice that $\alpha,\beta,\gamma$ violate
{\it both}  CPT, due to their decoherening nature~\cite{wald}, and
CP symmetry, as they do not commute with the CP operator
$\widehat{CP}$~\cite{lopez}: $\widehat{CP} = \sigma_3 \cos\theta +
\sigma_2 \sin\theta$,$~~~~~[\delta\H_{\alpha\beta}, \widehat{CP} ]
\ne 0$.
As pointed out in
\cite{benatti}, however,
in the case of $\phi$-factories complete positivity
is guaranteed within the above (single-particle) framework
only if the further conditions $\alpha = \gamma~ {\rm and} ~\beta = 0 $
are imposed.  Experimentally the complete positivity hypothesis,
and thus the above framework,
can be tested explicitly by keeping all three parameters.
In what follows, as far as single Kaon states are
concerned, we shall keep the $\alpha,\beta,\gamma$
parametrization~\cite{lopez},
and give the available experimental bounds for these parameters.
\begin{figure}[htb]
%\vspace*{9pt}
%\framebox[55mm]{\rule[-21mm]{0mm}{43mm}}
\centering
\epsfig{file=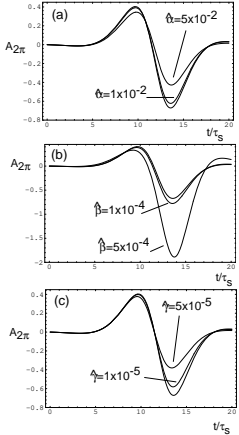, width=0.2\textwidth}
\vspace*{-0.5cm}\caption{Typical neutral kaon decay asymmetries
$A_{2\pi}$~\cite{lopez} indicating the effects of quantum-gravity
induced decoherence.} \label{AT}
\end{figure}
The relevant observables are defined as $ \VEV{O_i}= {\rm Tr}\,[O_i\rho] $. For
neutral kaons, one looks at decay asymmetries~\cite{lopez}
(c.f. fig.~\ref{AT} for the case of $2\pi$ final states). The
important point to notice is that the two types of CPTV, within
and outside
quantum mechanics, can be
{\it disentangled experimentally}~\cite{lopez}.

We next mention that, typically, for instance when the final states
are $2\pi$, one has  a time evolution of the decay rate $R_{2\pi}$:
$ R_{2\pi}(t)=c_S\, e^{-\Gamma_S t}+c_L\, e^{-\Gamma_L t} + 2c_I\,
e^{-\Gamma t}\cos(\Delta mt-\phi)$, where $S$=short-lived,
$L$=long-lived, $I$=interference term, $\Delta m = m_L - m_S$,
$\Delta \Gamma = \Gamma_S - \Gamma_L$, $\Gamma =\frac{1}{2}(\Gamma_S
+ \Gamma_L)$. One may thus define the {\it Decoherence Parameter}
$\zeta=1-{c_I\over\sqrt{c_Sc_L}}$, as a (phenomenological) measure
of quantum decoherence induced in the system. In our decoherence
scenario, $\zeta$ corresponds to a particular combination of the
decoherence parameters~\cite{lopez} $ \zeta \to \frac{\widehat
\gamma}{2|\epsilon ^2|} - 2\frac{{\widehat \beta}}{|\epsilon|}{\rm
sin} \phi~,$ with the notation $\widehat{\gamma} =\gamma/\Delta
\Gamma $, \emph{etc}.

The CPLEAR measurements gave the following bounds~\cite{cplear}
$\alpha < 4.0 \times 10^{-17} ~{\rm GeV}~, ~|\beta | < 2.3. \times
10^{-19} ~{\rm GeV}~, ~\gamma < 3.7 \times 10^{-21} ~{\rm GeV} $,
which are not much different from theoretically expected values in
some optimistic scenarios~\cite{lopez} $\alpha~,\beta~,\gamma =
O(\xi \frac{E^2}{M_{P}})$. The experiment KLOE at Da$\Phi$NE updated
these limits recently by measuring for the first time the $\gamma$
parameter for entangled Kaon states~\cite{adidomenico,testa}:
$\gamma_{\rm KLOE} = (1.1^{+2.9}_{-2.4} \pm 0.4) \times
10^{-21}~{\rm GeV}$, as well as the (naive) decoherence parameter
$\zeta$. This bound can be improved by an order of magnitude in
upgraded facilities, such as KLOE-2 at
DA$\Phi$NE-2~\cite{adidomenico}.

\vspace*{-0.3cm}
\section{CPTV and Modified EPR Correlations
of Entangled Neutral Kaon States} \vspace*{-0.0cm}

If CPT is \emph{intrinsically} violated,
in the sense of being not well-defined due to decoherence~\cite{wald},
the Neutral mesons $K^0$ and ${\overline K}^0$ should \emph{no
longer} be treated as \emph{indistinguishable particles}. As a
consequence~\cite{bmp}, the initial entangled state in $\phi$
factories $|i>$, after the $\phi$-meson decay, assumes the form:

{\scriptsize \begin{eqnarray*}
|i> &=& {\cal N} \bigg[ \left(|K_S({\vec
k}),K_L(-{\vec k})>
- |K_L({\vec k}),K_S(-{\vec k})> \right)\nonumber \\
 &+&  \omega \left(|K_S({\vec k}), K_S(-{\vec k})> - |K_L({\vec
k}),K_L(-{\vec k})> \right)  \bigg] \nonumber
\end{eqnarray*}}where $\omega = |\omega |e^{i\Omega}$ is a complex parameter,
parametrizing the intrinsic CPTV modifications of the EPR
correlations.
The $\omega$-parameter controls the amount of contamination of the
final (odd-under-permutation-symmetry(${\cal P}$)-) state 
by the ``wrong'' -symmetry 
(${\cal P}$(even)-) state.

\begin{figure}[htb]
\centering
  \epsfig{file=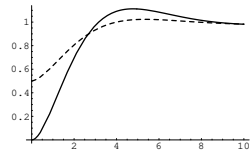, width=0.2\textwidth}
%\includegraphics[width=3.2cm]{fig2.eps}
%\hfill
%\includegraphics[width=3.2cm]{fig4.eps}
%\hfill\includegraphics[width=3.2cm]{fig1.eps} \hfill
%\includegraphics[width=3.2cm]{fig3.eps}
\vspace*{-0.3cm}\caption{A characteristic case of the intensity
$I(\Delta t)$, with $|\omega|=0$ (solid line)  vs  $I(\Delta t)$
(dashed line) with $|\omega|=|\eta_{+-}|$, $\Omega = \phi_{+-} -
0.16\pi$, for definiteness~\cite{bmp}.}
\label{intensomega}\end{figure}

The appropriate observable (c.f. fig.~\ref{intensomega})
is the ``intensity'' $I(\Delta t)
= \int_{\Delta t \equiv |t_1 - t_2|}^\infty
|A(X,Y)|^2$, with $A(X,Y)$ the appropriate $\phi$ decay
amplitude~\cite{bmp},
where one of the Kaon products decays to
the  final state $X$ at $t_1$ and the other to the final state $Y$
at time $t_2$ (with $t=0$ the moment of the $\phi$ decay).

The KLOE experiment has just released the first measurement of the
$\omega$ parameter~\cite{adidomenico,testa}: $ {\rm Re}(\omega) =
\left( 1.1^{+8.7}_{-5.3} \pm 0.9\right)\times 10^{-4}~$, ${\rm
Im}(\omega) = \left( 3.4^{+4.8}_{-5.0} \pm 0.6\right)\times
10^{-4}$. At least an order of magnitude improvement is expected
for upgraded facilities such as KLOE-2 at (the upgraded)
DA$\Phi$NE-2~\cite{adidomenico}. This sensitivity is not far from
certain optimistic models of space time foam leading to
$\omega$-like effects~\cite{bms}.

We close this section by mentioning that the $\omega$ effect can be
disentangled experimentally from \emph{both}, the C(even) background
- by means of different interference with the C(odd) resonant
contributions, and the decoherent evolution ($\alpha = \gamma$)
effects~\cite{bmp} - due to different structures. \vspace*{-0.3cm}
\section{Precision T, CP and CPT Tests with Charged Kaons}
\vspace*{-0.1cm} Precision tests of discrete symmetries can also be
performed with charged Kaons, which is a case that generated a great
interest in this conference~\cite{ali}, as a result of the (recently
acquired) high statistics at the NA48 experiment~\cite{NA48}, in
certain decay channels, which allows for precision tests of the
chiral perturbation theory~\cite{ali}. For our purposes of testing
CPT symmetry, we shall restrict ourselves to one particular charged
Kaon decay, $K^\pm \rightarrow \pi^+ + \pi^- + \ell ^\pm + \nu_\ell
(\overline \nu_\ell)$, abbreviated as $K_{\ell 4}^\pm$. One can
perform independent precision tests of T, CP and CPT using this
reaction~\cite{wu}, by comparing the decay rates of the $K^+$ mode with the
corresponding decays of the $K^-$ mode, as well as tests of $\Delta S =
\Delta Q$ and $|\Delta I|=1/2$ isospin rules. If CPT is violated,
through microscopic time irreversibility~\cite{wald}, then the phase
space analysis for the products of the reaction, from which one
obtains the di-pion strong-interaction phase shifts, needs to be
modified~\cite{wu}.

I would like to finish this section by mentioning the possibility of
exploiting the recently attained high statistics for charged Kaons
in the NA48 experiment~\cite{NA48} so as to use appropriate
combinations of \emph{both} reaction modes $K_{\ell 4}^\pm$ for
precision tests of physics beyond the Standard Model (SM), such as
supersymmetry, \emph{etc.}, including possible CPT violations. One
could look at T-odd triple momentum correlators~\cite{triple} $\vec
p_\ell \cdot (\vec p_{\pi_1} \times \vec p_{\pi_2})$. The so
constructed CP-violating observables are independent of the lepton
polarization and thus easier to measure in a high statistics
environment, such as the NA48 experiment~\cite{NA48}.
\vspace*{-0.3cm}
\section*{Acknowledgements}
\vspace*{-0.3cm} I thank the organizers and conveners of BEACH2006
Conference for the invitation. I
also thank J. Bernab\'eu, A. DiDomenico, J.
Ellis, D. Nanopoulos, J. Papavassiliou and Sarben Sarkar for discussions,
and B. Peyaud for an informative discussion on charged Kaon
decays.

\end{document}